# Synthesis and properties of free-standing monolayer amorphous carbon

Chee-Tat TOH*[1,2], Hongji ZHANG*[3], Junhao LIN[4,5], Alexander S. MAYOROV[2], Yun-Peng WANG[6, 7], Carlo OROFEO[1], Darim Badur FERRY[1], Henrik ANDERSEN[2], Nurbek KAKENOV[2], Zenglong GUO[5], Irfan Haider ABIDI[2], Hunter SIMS[6], Kazu SUENAGA[4,8], Sokrates T. PANTELIDES[6,9], Barbaros ÖZYILMAZ[1,2,3,†].

[1]Department of Physics, National University of Singapore, Singapore 117551, Singapore.

[2]Centre for Advanced 2D Materials, National University of Singapore, Singapore 117546, Singapore.

[3]Department of Materials Science and Engineering, National University of Singapore, Singapore 117575, Singapore.

[4]National Institute of Advanced Industrial Science and Technology, Tsukuba, Japan.

[5]Department of Physics, Southern University of Science and Technology, Shenzhen 518055, People's Republic of China.

[6]Department of Physics and Astronomy, Vanderbilt University, Nashville, Tennessee 37235, United States.

[7]Hunan Key Laboratory for Super-Microstructure and Ultrafast Process, School of Physics and Electronics, Central South University, Changsha 410083, Hunan, China.

[8]Department of Mechanical Engineering, The University of Tokyo, Japan.

[9]Department of Electrical Engineering and Computer Science, Vanderbilt University, Nashville, TN 37235, United States.

*These authors contributed equally to this work.

†Correspondence to phyob@nus.edu.sg

## Abstract

**Bulk amorphous materials have been studied extensively and are used widely. Yet, their atomic arrangement remains an open issue. They are generally believed to be Zachariasen continuous random networks (Z-CRNs)[1], but recent experimental evidence favours the competing crystallite model in the case of amorphous silicon[2–4]. Corresponding questions in 2D materials are wide open. Here we report the synthesis of centimetre-scale, freestanding, continuous, and stable monolayer amorphous carbon (MAC), topologically distinct from disordered graphene, by laser-assisted chemical vapour deposition[5]. Unlike bulk materials, the amorphous structure of MAC can be determined by atomic-resolution imaging. Extensive characterisation reveals complete absence of long-range periodicity and a threefold-coordinated structure with a wide distribution of bond lengths, bond angles, and 5-, 6-, 7-, and 8-member rings. The ring distribution is not a Z-CRN but resembles the competing (nano)crystallite model[6]. A corresponding model has been constructed and enables density-functional-theory calculations of MAC properties, in accord with observations. Direct measurements**

**confirm that it is insulating with resistivity values similar to CVD grown boron nitride. Freestanding MAC is surprisingly stable and deforms to a high breaking strength, without crack propagation from the point of fracture. The present work reveals a stable, freestanding MAC with excellent physical properties, potentially leading to unique applications.**

Main Text

Amorphous materials are used in a wide range of applications, but their atomic-scale structure and its effect on properties are far more complex than those of crystalline analogues. In a classic 1932 paper, Zachariasen[1] invoked free-energy arguments to propose that amorphous materials comprise the same bonding units as their crystalline analogues, but these units form continuous random networks (Z-CRNs) instead of periodic structures. A competing model, namely the existence of crystallites embedded in an otherwise CRN environment is even older[7] (the crystallite model of amorphous solids is distinct from nanocrystalline materials as the latter comprises nano-scale grains separated by grain boundaries). In the last few decades, the Z-CRN model gained wide acceptance, especially for amorphous Si and $SiO_2$ (a-Si and a-$SiO_2$), which are viewed as prototypes. The primary experimental evidence was provided by radial distribution functions (RDFs), which could be reproduced accurately by model Z-CRNs. More recently, however, experiments that probe beyond pair correlations found extensive concentrations of nanocrystallites of order 1-2 nm, while corresponding models fit the experimental RDFs just as well as Z-CRN models[2–4]. The inability to directly image the atomic-scale structure of bulk amorphous materials by any kind of microscopy, however, remains a limiting factor in the pursuit to resolve the CRN versus crystallite quandary[8].

For two-dimensional (2D), atomically thin materials, the nature of the amorphous state can in principle be resolved by direct atomic-resolution imaging. Monolayer amorphous carbon (MAC) can be viewed as a prototype amorphous 2D material, analogue of monolayer crystalline carbon, namely graphene. Atomic-resolution transmission electron microscopy (TEM) has been used to image freestanding graphene monolayers irradiated with the TEM electron beam to induce disorder[9,10]. Such monolayers, however, are limited in size, generally inhomogeneous because of carbon loss, and, when crystallites are still present, they are not necessarily randomly oriented, as they would be in a truly amorphous material, and can be eliminated by further irradiation. Similarly, radiation of aromatic self-assembled monolayers on substrates produces nanocrystalline graphene with only small amorphous patches[11].

Direct synthesis of MAC samples remains an open challenge despite prospects of unique and novel applications. Very recently, the synthesis of amorphous carbon monolayers on Ge substrates using conventional chemical vapour deposition (CVD) was reported[12]. These samples were grown at high temperatures (>900 °C) resulting in amorphous regions (~300 nm) that are embedded in a graphene matrix. The reported TEM images lack atomic resolution to establish the structure unambiguously. Instead, the authors concluded in favour of a Z-CRN structure on the basis of the halos in diffraction patterns derived from FFTs of TEM images of 2.5×2.5 nm$^2$ regions. However, many of the halos exhibit residual 6-dot motifs, i.e., the existence of crystallites 1 to 2 nm across cannot be ruled out.

In this paper, we report the synthesis of MAC by a laser-assisted chemical vapour deposition (CVD) growth process. The self-limiting process leads to a uniform monolayer of several square centimetres in less than one minute at substrate temperatures as low as 200 °C. Extensive atomic-resolution TEM characterisation shows unambiguously both the complete absence of long-range periodicity and a structure that is consistent with the crystallite model, i.e., nm-sized, randomly oriented, and strained crystallites comprising only 6-member rings embedded in a Z-CRN environment. For simplicity, here we discuss representative data on MAC films grown on a copper foil at 250 °C unless otherwise stated. Samples are easily transferred via wet etching after growth. Unlike other CVD-grown 2D materials, MAC is freestanding even on liquid surfaces without the need of a support polymer (Supplementary Video). Figures 1a and 1b show the transferred sample on a SiO$_2$/Si wafer and a TEM grid, respectively, to be homogeneous and continuous, free of multilayer regions or wrinkles that are typically observed with the transfer of 2D layers. In addition, MAC is stable and freestanding even when stored under ambient conditions for at least 1 year.

MAC samples were first characterised by Raman spectroscopy and X-ray Photoemission Spectroscopy (XPS) measurements. All samples have nearly identical spectra independent of substrate (Fig. 1c, d). The Raman 2D band at ~2680 cm$^{-1}$, which is pronounced in crystalline graphene, is negligible in the MAC, strongly suggesting the lack of any long-range order[13]. Furthermore, the Raman intensity ratio $I_D/I_G$ = 0.82 (Fig. 1e) indicates an average defect distance[14] of <1 nm. Note that Raman mapping of $I_D/I_G$ shows that MAC is completely uniform over areas of 50×50 μm$^2$ (Extended Data Fig. 1). Furthermore, C-1s X-ray photoelectron spectra (XPS) in Figure 1f show that the bonds in MAC are mainly C $sp^2$ (Supplementary Note 1).

Monochromated, aberration-corrected, high-resolution transmission electron microscopy (HRTEM) was used to image the exact arrangement of carbon atoms in MAC. A large-area HRTEM image reveals a connected but distorted structure of 5-, 6-, 7-, and 8-member rings (Fig. 2a). The Fourier transform in the inset of Figure 2a shows a broad, continuous halo. Figure 2b shows a zoomed-in 5×5 $nm^2$ area in which we can clearly see heavily distorted ~1 nm crystallites that are embedded in a CRN background that comprises 5-, 6-, -7, and 8-member rings. As the yellow arrows indicate, the crystallites are oriented randomly. This structure of MAC is consistent over the entire sample (Supplementary Note 2). Thus, MAC is topologically distinct from disordered graphene.

We also evaluated the pair correlation function (PCF) of neighbouring carbon atoms in MAC samples (Fig. 2e) and a graphene reference (Supplementary Fig. S8), a crucial property in deciding if a material is amorphous. Graphene peaks disappear or are highly broadened in the MAC PCF after the second nearest neighbours, confirming the loss of long-range periodic order. The bond lengths and angles for graphene are centred at 1.4 Å and 120°, respectively, with small variation from image aberration and algorithmic error. In contrast, MAC has much broader variation, 0.9-1.8 Å and 90-150° for the in-plane projections of bond lengths and bond angles, respectively (Fig. 2f, g). This feature is surprising, since the theoretical breaking strain at 25-30% on crystalline graphene occurs with deformations[15] at just under 1.6 Å and 135°, whereby one would expect freestanding MAC to be unstable and subject to collapse. On the other hand, such a wide spread of the bond angle distribution is expected for a random 2D amorphous network, similar to the O-Si-O bonds modelled in a silica bilayer on Ru[16]. Moreover, each carbon atom is threefold coordinated, consistent with the $sp^2$ bonding concluded from the XPS data.

Furthermore, the strained crystallites are fundamentally different from nanocrystalline graphene domains since they are not separated by atomically sharp grain boundaries but by CRN regions that are at least three carbon atoms wide (Fig. 2 h, i). Selective area electron diffraction (SAED) patterns confirm the amorphous nature of MAC by the characteristic diffuse halo (Fig. 2j), in contrast to sharp first and second order diffraction rings for nanocrystalline graphene (Fig. 2k). Dark-field TEM (DF-TEM) also reveals this, with only nanocrystalline graphene showing crystallinity (Extended Data Fig. 2 and Supplementary Note 3). To summarise, the direct imaging of the atomic structure clearly shows the amorphous nature of MAC to be (nano)crystallite, not Z-CRN.

The interlayer spacing of multilayer MAC was measured by atomic-force-microscopy (AFM) using overlapping MAC layers prepared by multiple wet transfers on a SiO$_2$/Si substrate. It was found to be ~0.6 nm (Fig. 3a and Extended Data Fig. 3). This interlayer spacing is almost twice as large as graphene's, similar to phosphorene's 0.6 nm, and consistent with the out-of-plane buckling caused by Stone-Wales defects[17]. AFM indentation experiments on suspended MAC membranes were used to determine the impact of structural buckling on its mechanical properties. The distribution of 2D elastic stiffness (E$_{2D}$) obtained by force-deflection measurements to 100 nN is centred at E$_{2D}$ = 115 Nm$^{-1}$ (Fig. 3c). We observed an increasing slope (hence stiffness) of the force-displacement curves with consecutive indentations of increasing force (Fig. 3d). Hence, MAC undergoes plastic deformation with applied force. Furthermore, we obtain a breaking strength of 22 Nm$^{-1}$, more than half the strength of single-crystal graphene[18]. We note that indentation rupture is restricted and does not propagate (Fig. 3c), in contrast to similar experiments with crystalline graphene which result in catastrophic failure by rapid crack propagation[18] (Supplementary Note 4).

Next, we discuss source-drain bias-dependent electronic transport measurements (I-V) as a function of temperature and layer number (see also Extended Data Fig. 4). In all samples we observed, $I = V^\alpha$, where $\alpha$ = ~2 (Fig. 4c) and sheet resistance values of order 100 GΩ at room temperature, similar to CVD BN[19]. In 2D, these values are among the most resistive and up to ~10 orders of magnitude larger than graphene. Furthermore, the large activation energy of 235 meV (obtained from the Arrhenius plot in Fig. 4c) is similar to what has been observed in CVD BN (193 meV)[20]. On the other hand, the temperature dependence of the resistivity exhibits (Fig. 4c) a power law dependence, $\rho = \alpha T^N$, instead of the expected Mott law, $\rho \propto e^{\left(\frac{T}{T_0}\right)^{-\gamma}}$, for variable range hopping (Extended Data Fig. 5). In addition, we observe that the temperature dependence of all I-V curves collapses to a single curve, indicating an apparent power-law dependence (Extended Data Fig. 6). The two results together suggest that charge transport in MAC is unique in 2D and mimics disordered quasi-one-dimensional (1D) systems known as rare-chain hopping (RCH)[21,22].

A periodically repeatable structural model of MAC was constructed to resemble the observed structure in Figure 2b, with similar ring size distribution and similar randomly oriented crystallites embedded in a Z-CRN network (Fig. 2c; see Extended Data Fig. 7 for the construction process). Relaxation of the structure using density-functional-theory (DFT) calculations produced buckling in the non-hexagon regions (Fig. 3a), as is known to occur at Stone-Wales defects in graphene[17]. We also constructed bilayer MAC by stacking the model structures at different relative displacements and minimised the total energy with respect to interlayer spacing. We obtained interlayer spacings ranging from 0.55 to 0.62 nm, in agreement with the measured interlayer spacings discussed above. Quantum molecular dynamics of the structure at 300 K reveal that the MAC is stable and undergoes out-of-plane distortions in the entire structure in the form of long-wavelength flexural phonons, similar to those in graphene[23].

The experimental results described above are in good agreement with electronic-structure and quantum-transport calculations performed using the MAC model described earlier (Fig. 2c). In Fig. 4e we show the calculated MAC density of states (DOS) and compare it with that of crystalline graphene. It is clear that the V-shaped DOS of crystalline graphene around the Dirac point is no longer present. Instead, the DOS exhibits a relatively broad peak at the nominal Fermi energy, suggesting an increase of localised states[24]. In Figure 4d, we plot the wave function of a state within this peak, showing significant localisation on non-hexagon rings that are separated by crystallites (Extended Data Fig. 8). The calculated transmission coefficient in the region around the Fermi energy, shown in the inset in Figure. 4e, approaches zero despite the high DOS, which explains the highly resistive charge transport.

To identify the effective band gap of MAC, we measured its optical properties. The optical transmittance of MAC is 98.1% at 550 nm and increases towards 99% in the infrared (Fig. 4b). Clearly, graphene's 2.3% minimum absorption defined by the fine structure constant is no longer applicable. More importantly, a Tauc plot gives an optical band gap of 2.1 eV with a long tail towards lower energies, which is reminiscent of "tail states" in 3D amorphous $Si$[25]. We also observed photoluminescence (PL) from MAC with a pronounced peak at 2.04 eV, which is not seen in graphene due to the absence of a bandgap. Charge localisation could be responsible for both the optical gap and the observation of photoluminescence, similar to insulating $sp^3$ amorphous carbon thin films with charge localisation at isolated $sp^2$ clusters[26]. Finally, ellipsometry confirmed a band gap at 2.1 eV and a corresponding relative permittivity of ~ 11 (Extended Data Fig. 9).

In summary, we demonstrated a method for the growth of large-area, freestanding monolayer of $sp^2$-bonded amorphous carbon films. This is the first example of a stable amorphous material in the 2D limit. We furthermore demonstrated that the ring distribution is not a Z-CRN but resembles the competing crystallite model. Overall, to date, there exists no solid evidence for the formation of homogeneous, stable, freestanding Z-CRN monolayers, though their possible existence cannot be ruled out. Finally, stable MAC films directly grown on a wide range of surfaces at low temperatures are likely to be useful for a wide range of applications, including anti-corrosion barriers for magnetic hard discs, for heat-assisted magnetic recording, current collectors and electrodes in batteries and supercapacitors. The possibility of engineering a distribution of carbon ring sizes also makes MAC attractive for selective ion-transport membranes for protons, lithium and other light ions.

**Main Figures**

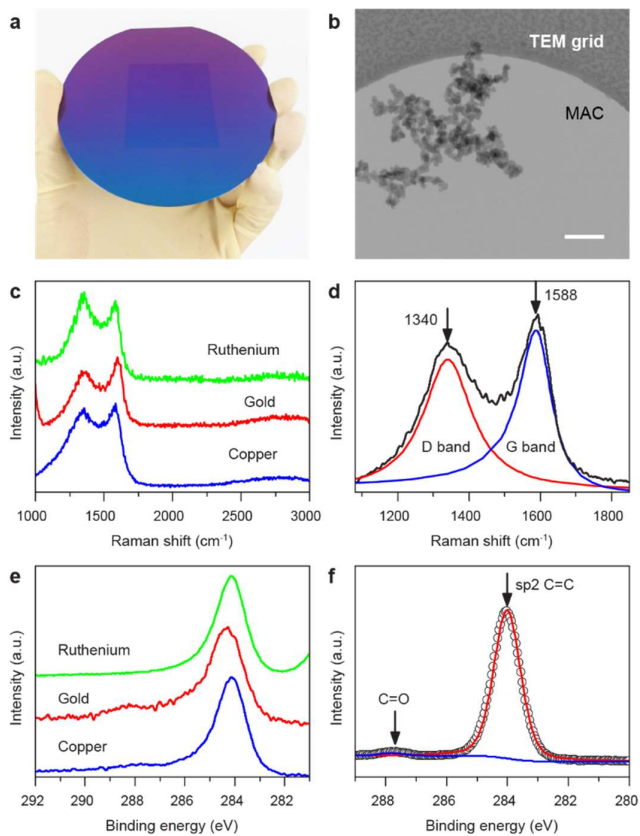

**Figure 1 Morphology of MAC. a**, 5×5 cm$^2$ of MAC transferred onto a SiO$_2$/Si wafer, and (**b**) SEM image of MAC suspended on a TEM grid has uniform contrast throughout. Nanoparticles intentionally suspended on MAC to emphasise the presence of the film. MAC grown on different substrates: **c**, Raman spectra for MAC grown on copper and gold substrate measured on after transferring to SiO$_2$, while MAC on ruthenium measured directly on the growth substrate. **d**, C-1s XPS spectra measured directly on different growth substrates. **e**, Raman spectra for growth on Cu showing the D and G bands with the fitted curve and an I(D)/I(G) ratio of 0.82. **f**, high-resolution C-1s XPS spectra on Cu with the fitted curve showing a single C sp$^2$ peak at 284.0 eV. Scale bar, 200 nm (**b**).

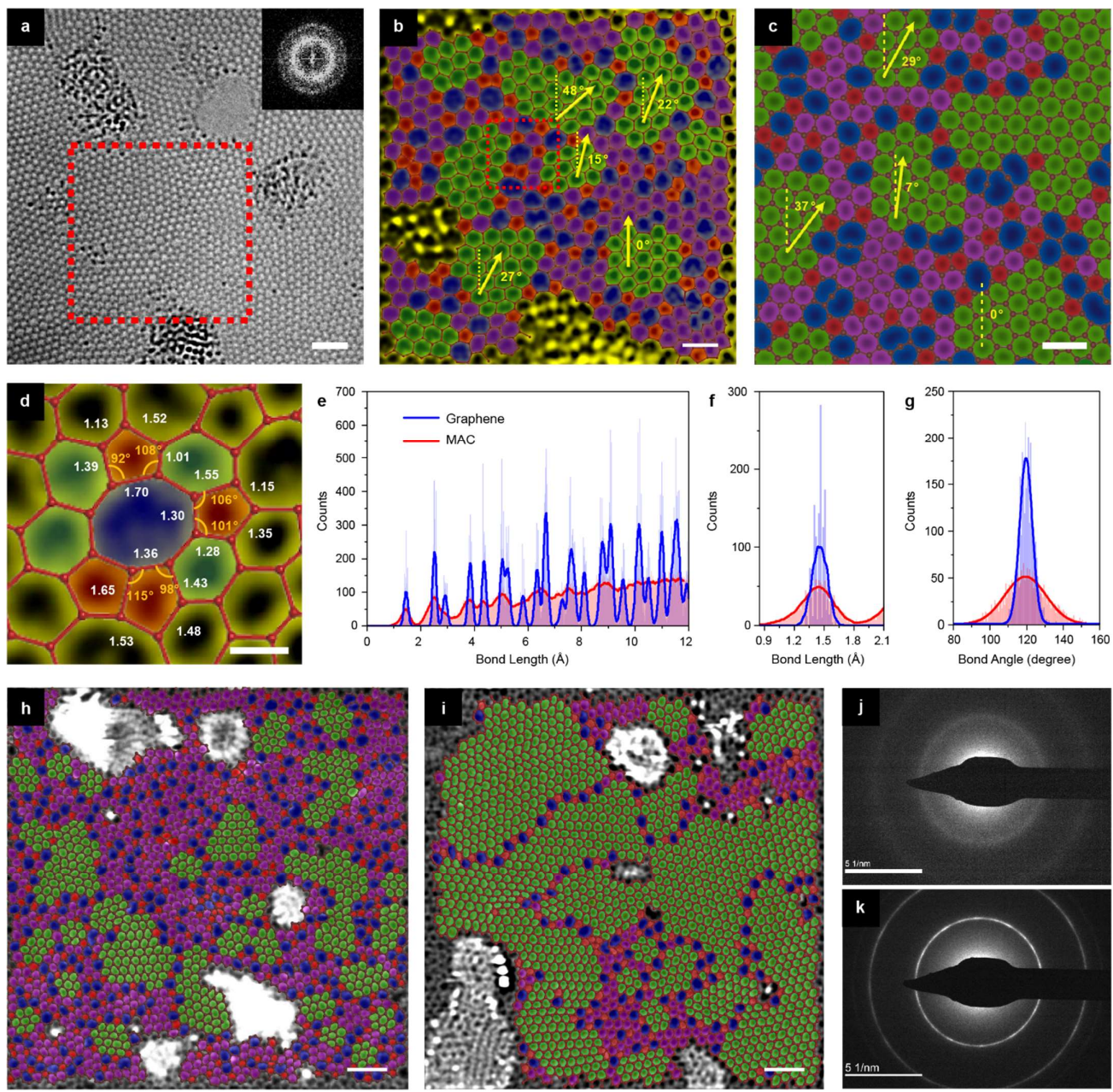

**Figure 2 Atomic structure of MAC from TEM. a**, 10×10 nm² monochromated HRTEM image of MAC and the corresponding Fourier transform of the selected region (inset). **b**, 5×5 nm² large scale atom-by-atom mapping of the selected region in (**a**). The contrast of the image is inverted and false coloured for better visibility. Colour overlay for identification of pentagons (red), heptagons/octagons (blue), strained hexagons (purple/green) that are omnipresent. Crystallites (green) separate regions with non-hexagons. Crystallites are defined to consist of at least a hexagonal ring surrounded by six hexagonal rings. Angular orientation of hexagons changes across the image, indicated by yellow arrows and the offset angle to the vertical. **c**, theoretical model created as described in Extended Data Fig. 7, replicating the MAC

features in (**b**), displayed with same colour coding. **d**, zoom-in region highlighted by red square in (**b**). The bond lengths (in Å) and bond angles of each pentagon are precisely measured. **e**, pair correlation function calculated by mapping coordinates of each carbon atom in (**b**). Graphene imaged under similar conditions with the same mapping algorithm is shown as a reference. **f**, statistical histogram of the bond length distribution to the first neighbouring atoms for MAC and graphene. **g**, statistical histogram of the bond angle distribution for MAC and graphene. STEM imaging comparison for **h**, MAC and **i**, nanocrystalline graphene with false colour overlay. Original images shown in Supplementary Figure S9. Selective area electron diffraction (SAED) patterns over a 3 μm diameter region of **j**, MAC sample shown in (**h**) and of **k**, nanocrystalline graphene sample shown in (**i**). Scale bars, 1 nm (**a**), 0.5 nm (**b**, **c**), 0.2 nm (**d**), 1 nm (**h, i**) and 5 1/nm (**j, k**).

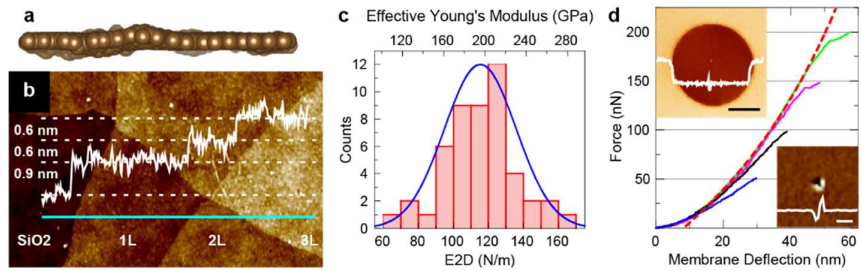

**Figure 3 Mechanical properties of MAC. a**, side view of the model MAC in Fig. 2c showing out-of-plane structural relaxation. **b**, AFM topography of 1-3 layers MAC on SiO$_2$/Si. Cyan solid line indicates line scan position for overlaid height profile. Image width, 3 μm. **c**, a histogram of the 2D elastic stiffness (E$_{2D}$) of 49 suspended MAC membranes with the applied force of 100 nN, with an average of E$_{2D}$ = 115 ± 21 Nm$^{-1}$. **d**, selected force-displacement curves from multiple indentations with increasing force at 25 nN steps. Fitting to linear elastic deformation expression indicated by the dotted red line. (top inset) AFM scan of MAC suspended on a 2.5 μm diameter well. White line is the height profile along the centre with 10.0 nm adhesion depth to the well wall. (bottom inset) Close up of collapsed MAC (White line is the height profile along the centre) after repeated indentations at 200 nN. Scale bars, 1 μm (**d**, top inset) and 100 nm (**d**, bottom inset).

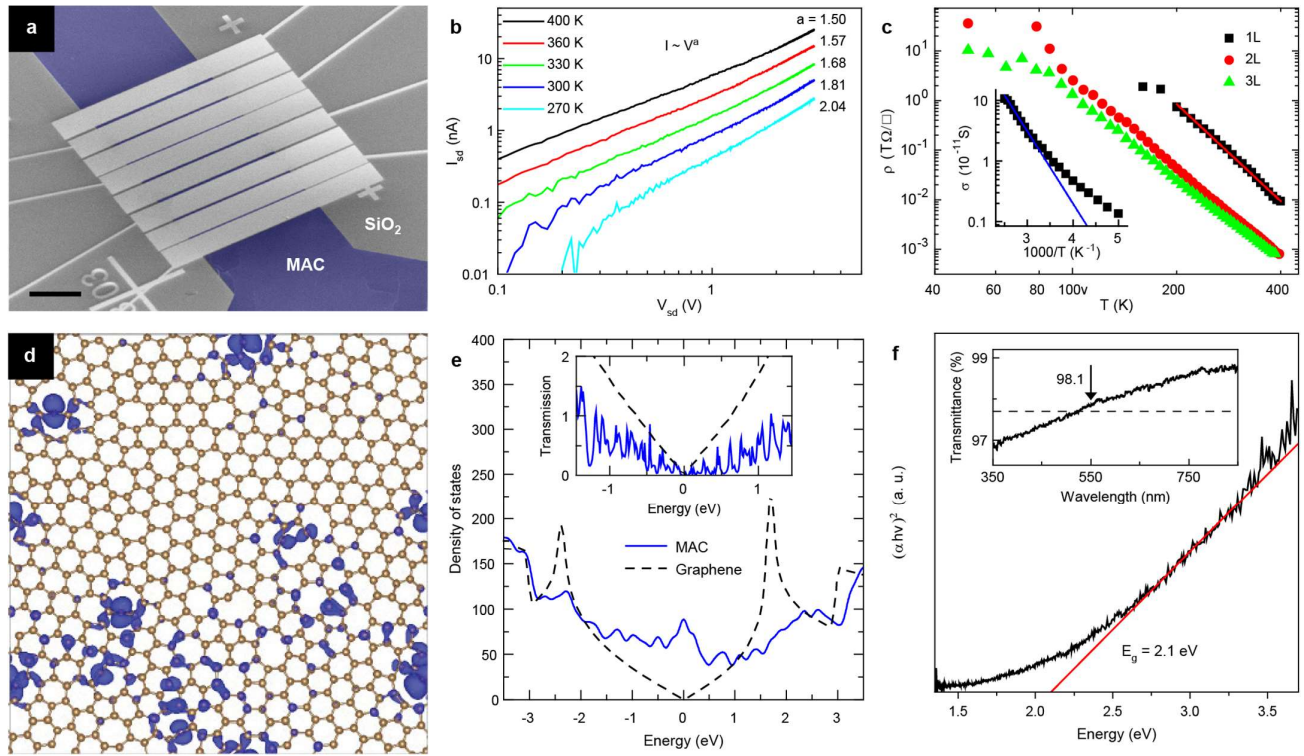

**Figure 4 Insulating properties of MAC. a,** false colour SEM image of the two-terminal device with 500 nm – 1 μm channel width and 50 μm long electrodes. **b**, nonlinear *I-V* curves measured at five different temperatures (Extended Data Fig. 4). **c**, resistivity for the samples with different layer numbers as a function of temperature. The red line is the linear best fit with the offset, *α*, and slope, *N*, equal to 10^(26.9±0.05) and -6.5±0.05, respectively. Leakage current from SiO$_2$/Si prevents lower temperature measurements. Inset: Arrhenius plot gives the activation energy of 235 meV. **d**, top view of the model MAC overlaid by the modulus square of the lowest unoccupied wave function. **e**, the density of states of the model MAC (solid blue line) shows a peak at the Fermi energy compared with that of graphene (dashed black line). Inset: phase-coherent electronic transmission through the model MAC (solid blue line) as compared with that through graphene (dashed black line). **f**, Tauc plot to determine the optical bandgap in amorphous materials, with extrapolation of the linear region (red line) estimating an optical band gap of 2.1 eV. Inset: optical transmittance of MAC on CaF$_2$ substrate. The dashed line indicates pristine graphene absorption at 2.3%. Scale bar, 20 μm (**a**).

## Methods

### MAC Synthesis

MAC was synthesised by laser-assisted chemical vapour deposition (LCVD) process with Krypton Fluoride (KrF) pulsed laser ($\lambda$= 248 nm, Coherent Inc.). Cu foils with 35 µm thickness were cleaned by sonication in acetone and isopropyl, followed by 30 min annealing in $H_2$ at 1,010°C in a quartz tube furnace. Sputtered Au and Ru substrates were used without prior annealing. The growth substrates were transferred into the LCVD vacuum chamber and mounted on the titanium stage. The chamber was evacuated to base pressure of $10^{-6}$ mbar. $CH_4$ was introduced into the chamber to $10^{-2}$ mbar pressure. The substrate was exposed to a plasma (350 kHz pulsed DC generator at 20 W) and pulsed laser (40 mJ/cm$^2$, 50 Hz) over a 5 cm$^2$ area, for complete film formation under 1 min. To obtain larger coverage, the stage was rastered with constant motion relative to the laser spot. Substrate temperatures were monitored by an IR thermal sensor (Optis® CTlaser 3MH$_2$-CF$_4$). Growth durations ranged from 1 min to 10 mins, with temperatures due to laser heating between 200 °C to 500 °C; MAC growth was of similar quality. Substrate temperatures can be further controlled by a stage heater, for nanocrystalline graphene grown between 500 °C to 650 °C, and polycrystalline graphene grown at >650 °C. While MAC synthesised on both sides of the copper foil has similar amorphous quality, MAC from the underside (facing away from the laser) was used for detailed characterisations.

### MAC sample preparation

MAC from the top side of the substrate was first removed by 5 mins of 50 W Ar plasma. The Cu foil was wet etched in 0.7 wt% Ammonium Persulphate solution $(NH_4)_2S_2O_8$, by floating the growth substrate on the solution, with MAC facing up without a polymer support. It was transferred to DI water for 2 hours twice using a rigid substrate to rinse the interface before scooping onto the desired substrate (SiO$_2$/Si wafers or PELCO Holey SiN TEM grids with 2.5 µm pores). For transfer of MAC to SiO$_2$/Si substrate with holes for indentation experiment, spin coated PMMA (495 PMMA, A4, 4% in anisole, 4,000 rpm) was required as a polymer support during the transfer process. Spin coated samples were baked on a hotplate at 180 °C for 2 minutes after spin-coating and the same transfer steps were performed. Suspended MAC were dried supercritically from acetone in a critical point dryer (CPD) system. The samples were annealed in 5% $H_2$ in Ar gas for 5 hours for removal of any organic residues.

**Sample Characterisation**

SEM images were acquired using FEI Verios 460 Field Emission Scanning Microscope. Raman spectroscopy was measured by ALPHA 300 R from WITEC (532 nm excitation laser, 1 µm diameter circular laser spot). Raman spectrums for the MAC grown on Cu and Au were measured after transfer to $SiO_2$ substrate while Raman spectrum was directly obtained on Ru substrate. XPS spectra were measured using UHV Vacuum Generators ESCALAB Mk2 system with Omicron 7-Channeltron analyser, directly on the growth substrate for Cu, Au and Ru. High-resolution Angle Resolved XPS spectrum on Cu was measured using the Singapore Synchrotron Light Source at National University of Singapore with Omicron EA 125 hemisphere energy analyser. XPS C1s peaks were fitted after a Shirley background subtraction using XPSPEAK software with a 30% of Lorentzian-Gaussian ratio. The fitted data show a single $sp^2$-hybridised carbon peak with full width half maximum of 1.0. MAC was transferred to Calcium Difluoride substrate for optical measurements. The UV–vis absorbance was recorded at room temperature on a SHIMADZU UV-3600 spectrophotometer. Photoluminescence (PL) measurements were performed with an NT-MDT NTEGRA SPECTRA confocal Raman microscope in backscattering geometry. We used the excitation laser at 473 nm and a 100× objective lens with numerical aperture of 0.6 to focus onto ~1.5 µm spot size. The laser density of power was kept below 1 $kW/cm^2$ in order to avoid any possible heating effects. The PL signal was detected with a CCD array operating at -80°C. Optical measurements were performed in high vacuum ($10^{-5}$ mbar).

**TEM**

The DF-TEM imaging was conducted in a FEI Tecnai F30 microscope operating at 80 kV. The DF-TEM images were recorded using the diffraction rings selected by the objective aperture, with a recording time of 120 s for each image. MAC HRTEM imaging was performed on a non-commercialised JEOL ARM60 microscope equipped with a Schottky field emission gun, a JEOL double Wien filter monochromator, double delta corrector and a Gatan one view camera with high stability. A slit of 2.8 µm is used for energy filtering, enabling a spatial resolution of 1.1 Å with a beam current of ~20 pA. The microscope was operated at 60 kV. The STEM imaging was done in a JEOL 2100F with delta probe corrector, which corrects the aberration up to 5th order, resulting in a probe size of 1.0 Å. The imaging was conducted at an acceleration voltage of 60 kV. The convergent angle for illumination is about 35 mrad, with a collection detector angle ranging from 45 to 200 mrad. A JEOL heating-holder is used in all experiments to heat the sample up to 700 °C during imaging. The pair correlation function and bond angle distribution are calculated by the coordinates of the carbon atoms determined in the HRTEM and STEM images by a home-built atom-finding algorithm.

**Computational Method**

The theoretical model for the crystallite MAC was constructed using the kinetic Monte Carlo method[27]. The simulation starts with a 40×40 Å² supercell containing 610 carbon atoms randomly placed in the x-y plane. The density of carbon atoms is the same as crystalline graphene. The supercell is then relaxed using a conjugated gradient algorithm. The inter-atom interactions are described using the adaptive intermolecular reactive empirical bond order (AIREBO) potential by Stuart[28] as implemented in LAMMPS[29]. Starting from the initial configuration, kinetic Monte Carlo is used for annealing. In each Monte Carlo iteration, one Stone-Wales transformation is performed to the current configuration, that is, one randomly chosen bonded carbon-carbon pair is rotated by 90 degrees in the x-y plane. The relaxed new structure is accepted by a probability min(1, exp(($E_{old}$-$E_{new}$)/$k_B$T)), where $E_{old}$ is the energy of current configuration, $E_{new}$ is the energy of the new configuration after relaxation, $k_B$T is selected to be 0.5 eV. We perform more than 60,000 iterations and generate 1,560 distinct configurations. We select the configuration at the 20,000$^{th}$ iteration as the theoretical model for its similarity with the experimental AC-HRTEM image (Extended Data Fig. 7). The electronic structure is calculated using density functional theory in the generalised gradient approximation[30] as implemented in the VASP software[31]. The coherent electronic transmission probability is calculated using the Transiesta program.

**Mechanical characterisation**

AFM topography was measured by tapping mode using Bruker AFM. This was also used for indentation experiments. MAC was suspended on wells with diameter of 1 to 2.5 μm. Force displacement curves were obtained by the indentation experiments performed at the centre of the suspended membranes using single crystal diamond probes with 8 nm diamond tips (spring constant of 1-5 Nm$^{-1}$) on silicon cantilevers, manufactured by K-TEK nanotechnology. The applied force was then increased till fracture to obtain the breaking strength of MAC. The 2D Elastic constant of MAC was calculated by fitting the force-displacement using the equation for force[32],

$$F = \frac{2\pi\sigma\delta}{\ln\left(\frac{a}{r}\right)} + \frac{Eq^3\delta^3}{a^2} \qquad (1)$$

where $a$ and $r$ are the radius of the suspended membrane and the AFM tip, respectively. $\sigma, \delta$ and $E$ are the 2D pretension, deflection and 2D effective Young's modulus of the membrane, respectively. $q$ (~ 1.02) is a function of Poisson's ratio (taken as 0.165).

The breaking strength corresponding to a breaking force $F_b$ was then calculated using,

$$\sigma^{2D} = \left(\frac{F_b E}{4\pi r}\right)^{\frac{1}{2}} \qquad (2)$$

**Electronic transport characterization**

Multi-terminal devices of one-, two-, and three-layer-thick MAC were fabricated on a Si/SiO$_2$ wafer using the standard microfabrication procedures of electron beam lithography. Oxygen plasma was first used to pattern MAC to fixed width of 50 µm. Devices were created with channel length between 200 nm and 1,000 nm. The width-to-length ratio ranged from 50 to 250. The geometry of each sample is shown in Supplementary Table. After the second step of lithography, a set of Co/MgO 35 nm/7 nm thick electrodes were deposited in an ultra-high vacuum system by an e-beam evaporator at the deposition pressure of 1e$^{-8}$ Torr. 10 µm wide contacts were made to reduce the contact resistance (Fig. 4a). Electrical measurements were carried out in the DC regime using two Keithley 2400 Sourcemeters for source-drain and gate voltages, which allows applying a voltage and measure the current simultaneously. The gate leakage was lower than 1 nA at room temperature and below 25 pA at 150 K. All measurements, for 2D amorphous carbon films, were performed in helium-free cryostats and in vacuum better than 10$^{-5}$ mbar to ensure reproducible measurement conditions. Resistivity measurements are performed at source-drain voltage of 1 V. Resistance was extracted from the I-V characteristic using a polynomial fit and taken at the first derivative at zero voltage.

**Methods References**

**Extended Data Figure Legends**

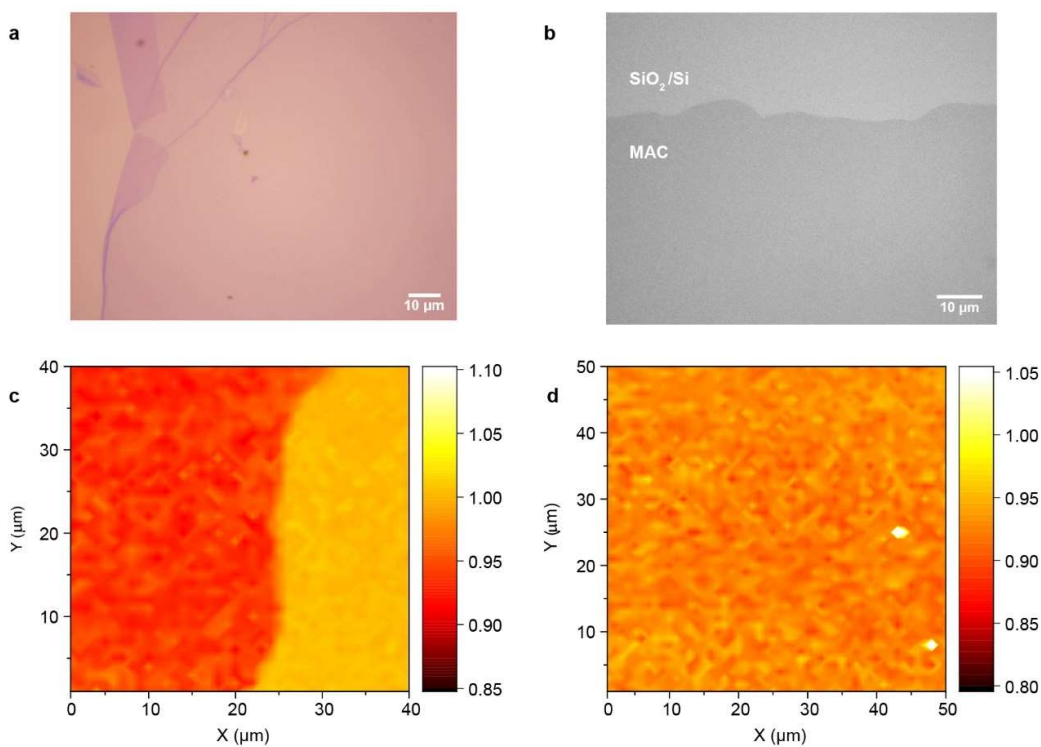

**Extended Data Fig. 1 Optical image and Raman mapping of MAC film transferred on Si/SiO2 substrate. a**, MAC transferred to SiO2/Si wafer. Crease and fold at MAC edge from transfer process. **b**, contrast enhanced image in greyscale shows no topological features visible on MAC. **c**, Raman mapping shows the intensity ratio of the D to G band, I(D)/I(G), is uniform at the edge of transferred MAC on a Si/SiO$_2$ substrate. This ratio is slightly higher than the expect ratio of 0.82, since Raman background signal is not subtracted during Raman mapping. **d**, 50×50 μm$^2$ Raman map (with 2,500 data points) of I(D)/I(G) shows uniform quality over large area (Standard deviation is ±1.8%).

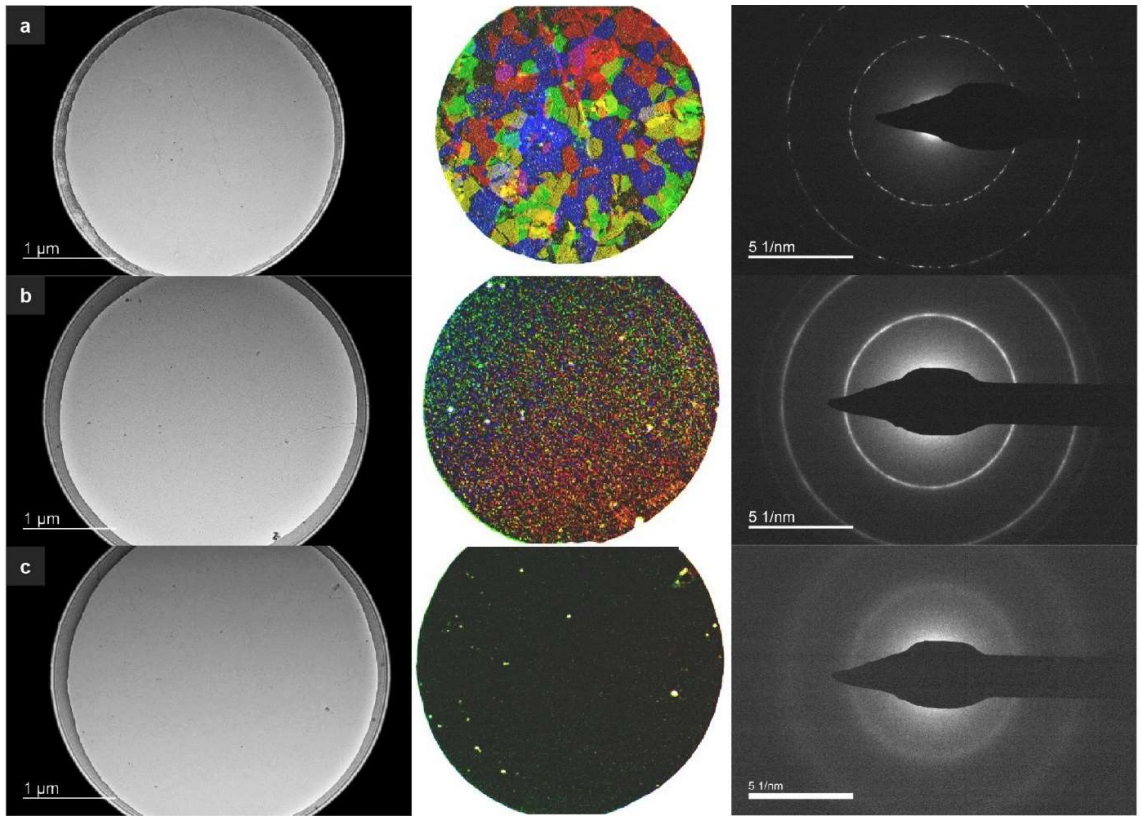

**Extended Data Fig. 2 Comparison of large-area TEM data**. **Row a,** polycrystalline graphene with 200-500 nm grains, **Row b,** nanocrystalline graphene (sample in Fig. 2i) with 1-3 nm grains (Row b) and, **Row c,** MAC (sample in Fig. 2h). **Left Column**: Bright-field TEM (BF-TEM), **Centre Column**: dark-field TEM (DF-TEM) with false colour image overlay showing crystal domains, and **Right Column**: Selective Area Electron Diffraction (SAED) patterns. Both polycrystalline and nanocrystalline graphene show crystal domains under DF-TEM and have well-defined SAED diffraction patterns. MAC has no visible domains in DF-TEM. Only MAC has the characteristic amorphous halo from SAED diffraction.

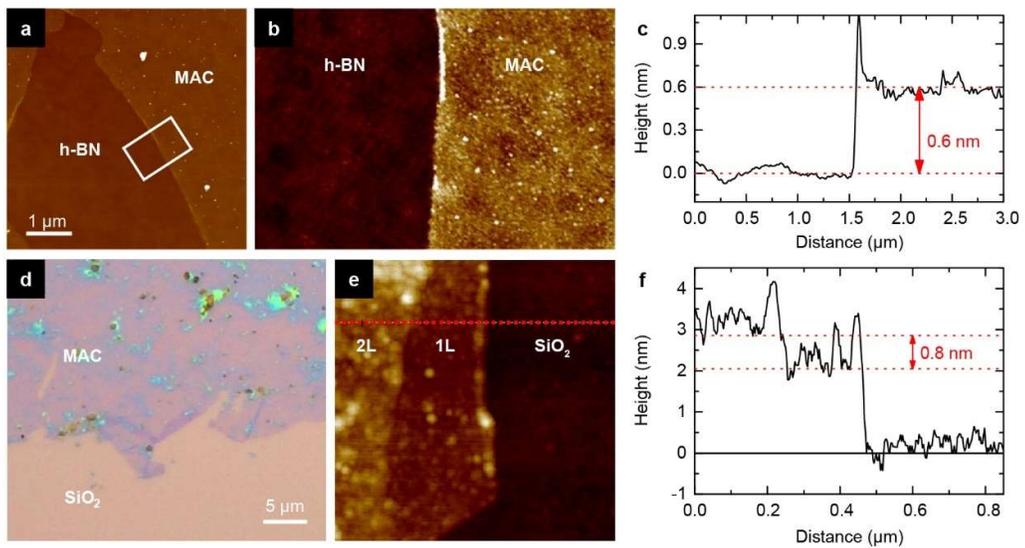

**Extended Data Fig. 3 Thickness of MAC. a**, AFM topography of MAC transferred on atomically flat, exfoliated BN crystal. **b**, close up of region in (**a**) indicated by white box with scan length of 3 μm. **c**, height profile of MAC obtained by averaging vertically over entire image scan in (**b**) shows a thickness of ~0.6 nm. **d**, optical image of MAC grown on gold substrates and transferred onto $SiO_2$ by wet transfer method with $KI/I_2$ solution as gold etchant and followed by 7 hours annealing in $Ar/H_2$ at 300°C. For MAC transferred from Au, residues from the gold etching cannot be fully removed due to non-optimized transfer process. **e**, AFM topography of a folded MAC edge in (**d**) to obtain 1 and 2 layers. **f**, height profile along the red dotted line in (**e**) give the thickness of MAC grown on gold be ~0.8 nm. Residual contamination from the transfer step and difference in substrate interactions may account for the higher thickness of MAC when transferred from Au.

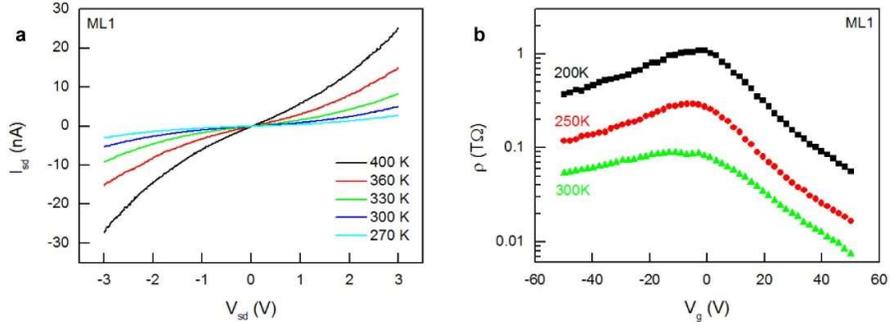

**Extended Data Fig. 4 IV measurement and gate dependence. a**, nonlinear *I-V* curves in Fig. 4b shown in linear scale. **b**, Gate dependence of MAC with electrostatic gating by $SiO_2/Si$ back gate shows ambipolar behaviour and strong temperature dependence, increasing from 100 GΩ/□ at room temperature to 1TΩ/□ at 200K.

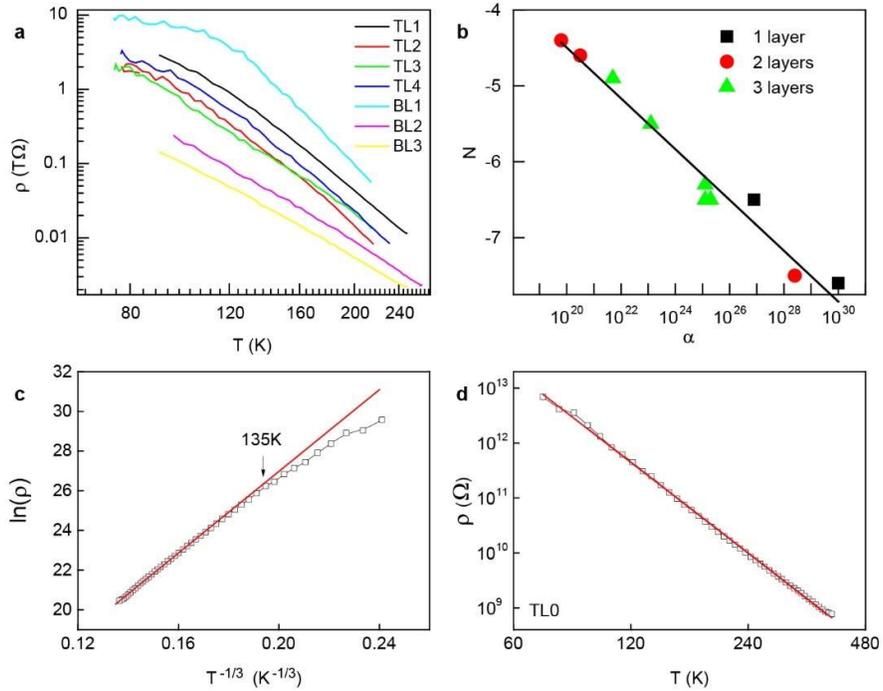

**Extended Data Fig. 5 Resistivity of samples as a function of the temperature. a**, the temperature dependence of the resistivity for a set of 2- and 3-layer devices (denoted with BL and TL respectively). **b**, Linear fit of resistivity of offset, *α*, and the power, *N*, as described by the formula $\rho = \alpha T^N$, shows a linear correlation, $N =$

2.2(±0.6) - 0.33(±0.02)log$_{10}α$. The data in (**a, b**) is provided in Supplementary Table. **c-d**, resistivity of tri-layer amorphous carbon sample (TL0) as a function of temperature plotted in linear scale. The Mott plot (**c**) shows a temperature limit (indicated by the black arrow) where it can be applied in comparison with the power-law dependence (**d**). The red lines are the best linear fits.

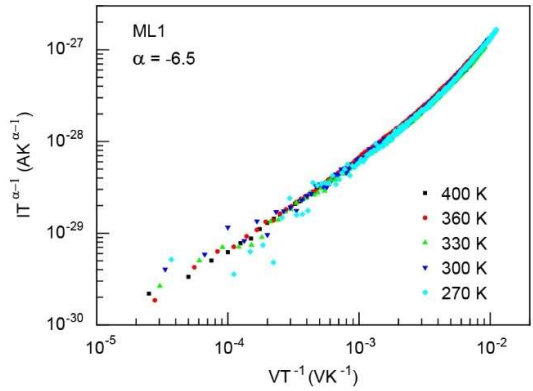

**Extended Data Fig. 6 Universal scaling of non-linear I-V characteristics for ML1 sample.** All I-Vs are collapsed to a single curve indicating apparent power-law dependence in disordered quasi-one-dimensional system. Equation for the universal scaling curve obtained from Rodin and Fogler[21].

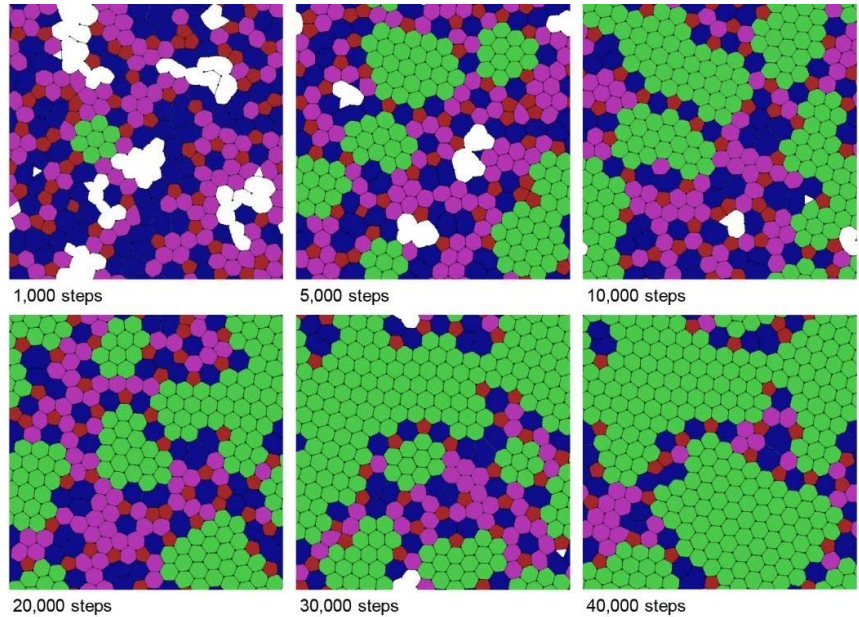

**Extended Data Fig. 7 Selection of MAC theoretical model.** Configuration for MAC structure at the 20,000th iteration is chosen as the theoretical model, shown in Fig. 2c, for its similarity with experimental AC-HRTEM images. Previous steps also show unstable ring structure (e.g. <4-carbon atom rings or ill-defined rings). Furthermore, ring size distribution is similar from 10,000 steps. Colour overlay for identification of pentagons (red), heptagons/octagons (blue), strained hexagons (purple/green) that are omnipresent. Crystallites (green) separate regions with non-hexagons. Crystallites are defined to consist of at least a hexagonal ring surrounded by six hexagonal rings.

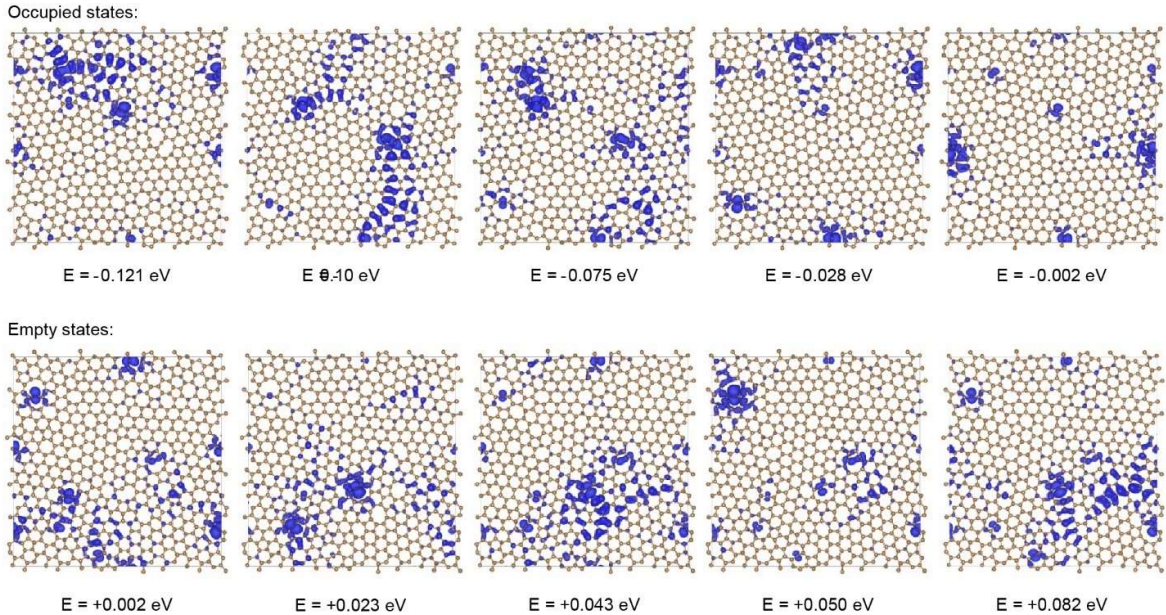

**Extended Data Fig. 8 Modulus square of electronic wave functions.** The lowest unoccupied wave functions are shown for ten states with energies closest to the Fermi energy. This shows significant localisation on non-hexagon rings that are separated by crystallites.

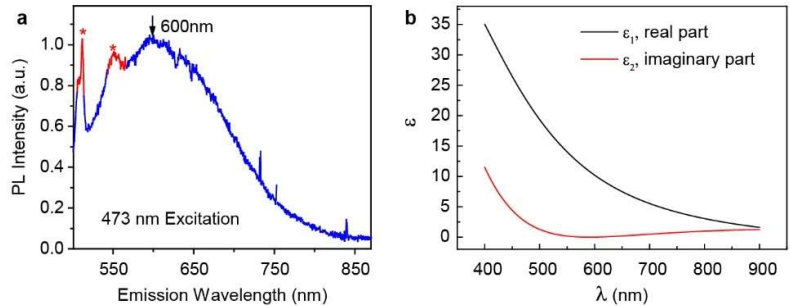

**Extended Data Fig. 9 Optical measurements of MAC on SiO$_2$/Si. a**, photoluminescence spectra was measured with excitation wavelength at 473 nm. The peaks marked by asterisks correspond to the Raman signal from MAC. The emission spectrum of the film has a broad peak centred at 600 nm. **b**, relative permittivity of MAC on SiO$_2$/Si substrate by ellipsometry measurements. $E_g$ is taken where imaginary part equals zero at ~590 nm (2.1 eV), at which the real part of the relative permittivity is ~11.

# Supplementary Information: Synthesis and properties of free-standing monolayer amorphous carbon

Chee-Tat Toh, Hongji Zhang, Junhao Lin, Alexander S. Mayorov, Yun-Peng Wang, Carlo Orofeo, Darim Badur Ferry, Henrik Anderson, Nurbek Kakenov, Zenglong Guo, Irfan Haider Abidi, Hunter Sims, Kazu Suenaga, Sokrates T. Pantelides, Barbaros Özyilmaz.

**Supplementary Note 1: XPS Analysis of MAC**

MAC synthesised on various substrates were analysed by Angle Resolved X-ray Photoemission Spectroscopy (XPS) to identify the chemical state and elemental composition. The fitted peaks in Fig. 1f were assigned to sp$^2$ hybridised C=C bonds and C=O bonds. Here, we established that C=O bonds is attached to the Cu substrate and does not contribute to the MAC structure.

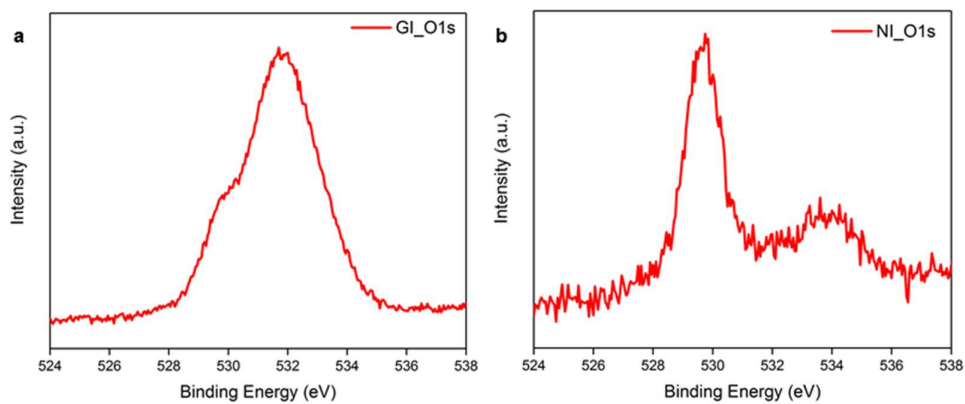

**Figure S1 O1s XPS spectrum for MAC on copper. a**, spectra for grazing incidence (GI) orientation which has high surface sensitivity. Dominant peak at 532 eV for CO group, peak at 529.5 eV for CuO **b**, spectra for normal incidence (NI) orientation with dominant signal from CuO.

C-1s XPS spectra estimated 2% of CO bonds from the fitting at 287.8 eV. O1s spectrum was measured to identify the source of oxygen content, as shown in Fig. S1. Two orientations were used during the measurement, with the grazing incidence (GI) orientation (Fig. S1a) providing a surface enhanced signal over the normal incidence (NI) orientation (Fig. S1b). GI has a dominant peak at 532 eV for CO group, and an additional peak at 529.5 eV for CuO. NI has a single peak at 529.5 eV for CuO. The synthesis process occurs in a room-temperature environment, with unavoidable high moisture content in the vacuum chamber and high oxide content in the Cu growth substrate. Interaction between C and CuO leads to the formation of copper carbonyl complexes on the surface, typically observed by the surface specific CO group and CuO. This suggest that C=O bonds observed in the C-1s spectra belongs to the CO group adsorbed on Cu and is not part of MAC's structure.

**Supplementary Note 2: Large Area Uniformity of MAC**

We investigated continuous MAC films of 4×4 cm² which were wet etch from copper and transferred from the etchant and water surface without a polymer support (Fig. S2). This feature is crucial for reducing transfer-related defects and contamination. In contrast, graphene collapses without polymer support during wet transfer under the surface tension of water[33].

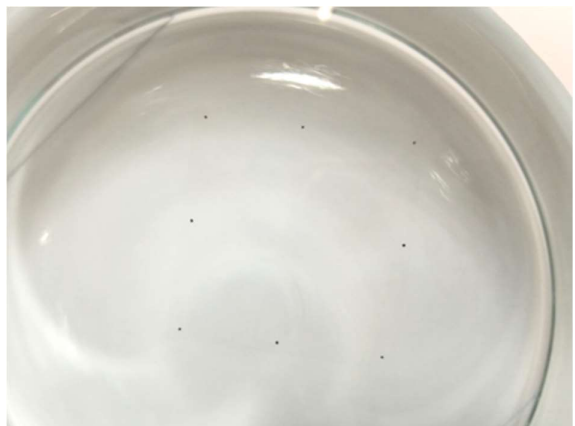

**Figure S2 Transfer of MAC.** 4×4 cm², unsupported monoatomic layer of carbon (MAC) floating on the surface of water. For visibility, black dots were marked on the corners and edge of the sheet before removal of growth substrate. Video of MAC moving around the water surface is available in Supplementary Video.

For atomic resolution TEM imaging, we avoid contributions from hydrocarbon absorbents and prevent carbon re-deposition by following standard precaution and imaging at elevated temperatures. TEM images of MAC structure at 450 °C and 700 °C are similar. This thermal stability is consistent with the Raman spectra of MAC obtained after thermal annealing at 700 °C (Fig. S3). Therefore, we use 700 °C to minimise residual contamination to typically around 10% surface coverage and includes Si atoms.

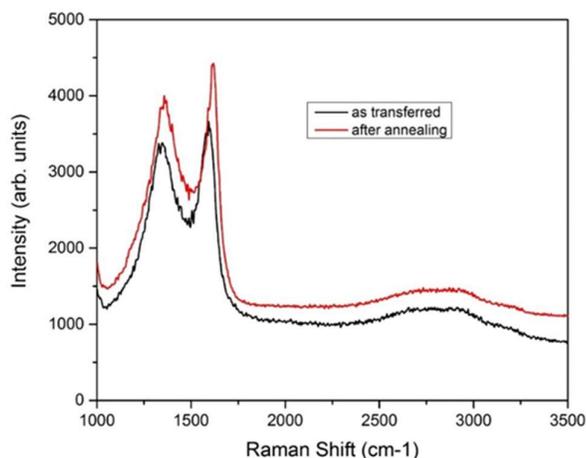

**Figure S3 Stability of MAC.** Raman spectra of MAC on SiO$_2$/Si before and after annealing in vacuum to 700 °C. No change to $I_D/I_G$.

MAC from random locations were transferred on 3 mm diameter TEM grids and more than 100 TEM images were taken. SEM of suspended MAC is consistently uniform over all images and representative images are shown in Figure S4. This is different from graphene, where grain boundaries and multilayer islands are visible.

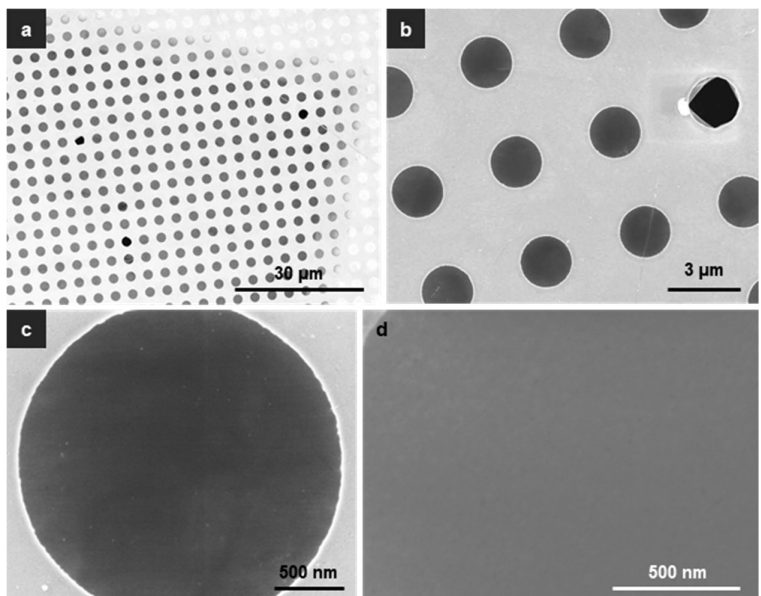

**Figure S4 SEM images of MAC, a-c**, on a TEM grids with field of view from ~100 μm down to 1 μm. **d**, zoom in to the edge of an individual well showing a uniform sheet free of residues.

Similarly, TEM images of MAC is consistently uniform and representative images are shown for TEM diffraction patterns taken from spots separated by 2-10 μm (Fig. S5), atomic-resolution TEM images selected from a 70×70 nm$^2$ area (Fig. S6), and high-resolution STEM images from areas separated by at least 100 μm (Fig. S7), showing the uniformity of our samples.

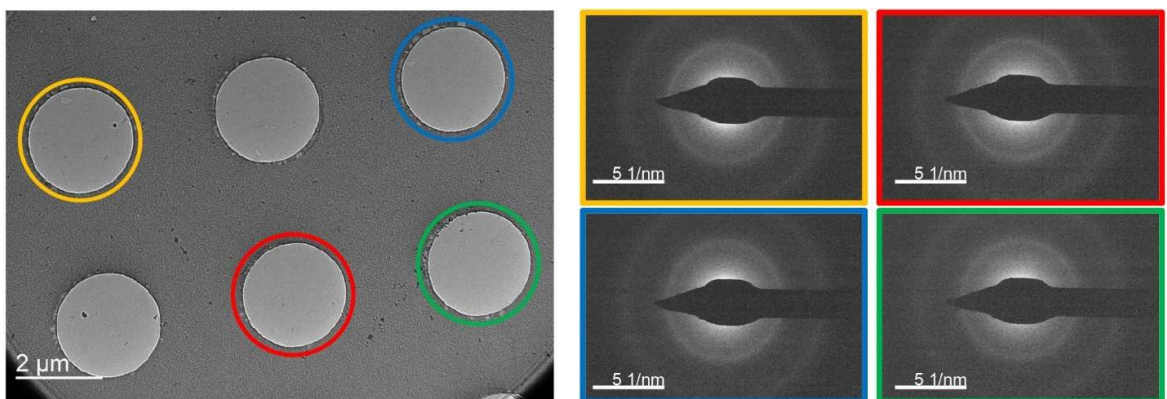

**Figure S5 TEM diffraction of MAC**. Diffraction was measured over 4 different areas (colours indicate diffraction measured at corresponding location) separated by up to 10 microns. They show identical diffraction patterns. MAC is therefore uniform over the micron scale.

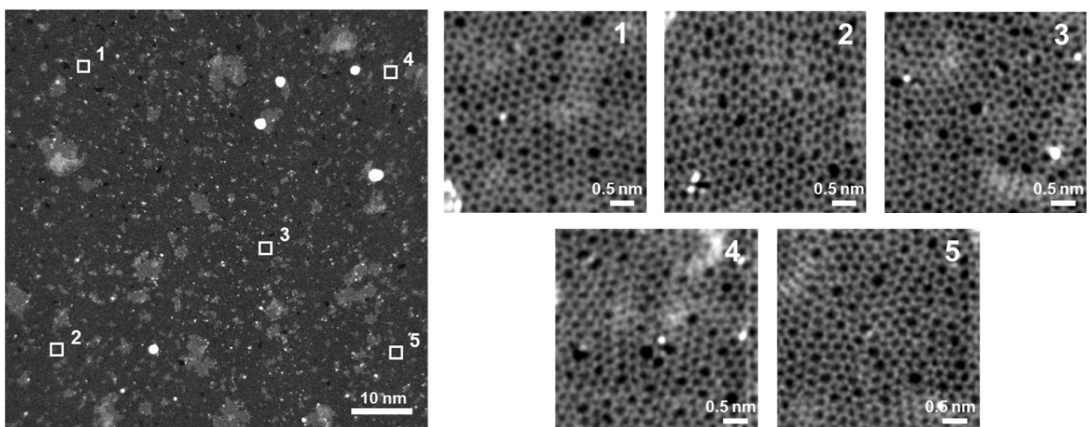

**Figure S6 Atomic resolution STEM images of MAC.** Images 1-5 are atomic resolution STEM images of MAC at the selected areas marked on the corresponding 70×70 nm$^2$ STEM image.

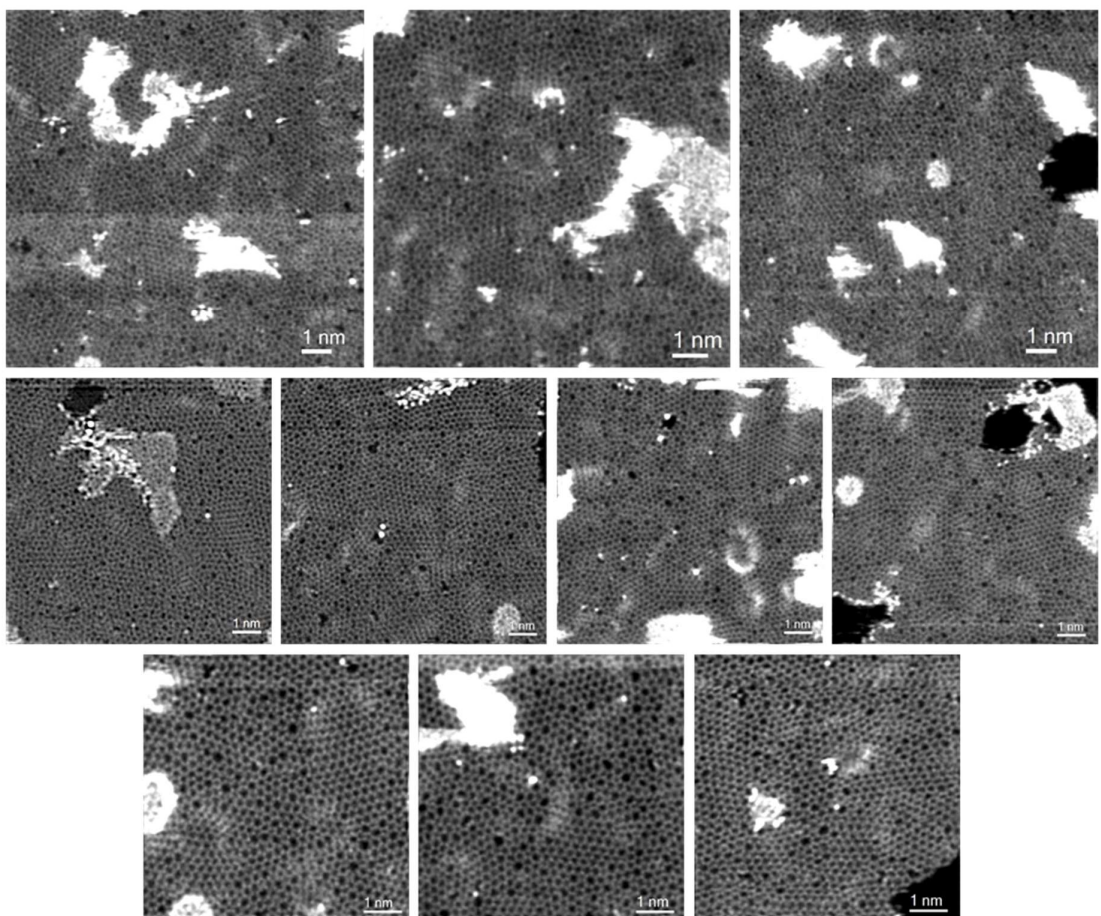

**Figure S7 High resolution STEM images of MAC** Randomly selected sites at least 100 μm apart from each other were imaged across the TEM grid at different magnifications (7×7 nm$^2$ to 12×12 nm$^2$). This shows that the amorphous structure of MAC as discussed in the main text is uniform over the entire TEM grid (~1 mm$^2$).

**Supplementary Note 3: Fundamental Differences of MAC to Graphene**

MAC meets the requirement of an amorphous material from the evaluation of the pair correlation function (PCF). The loss of long-range periodic order in the PCF is because of heavily strained bonds in MAC and this contrasts with crystalline graphene with sharp peaks well beyond the second nearest neighbour (Fig. 2e). This distribution of bond length and bond angle variation is clear when comparing crystalline graphene hexagonal lattice to the wide distribution of carbon ring sizes for MAC (Fig. S8).

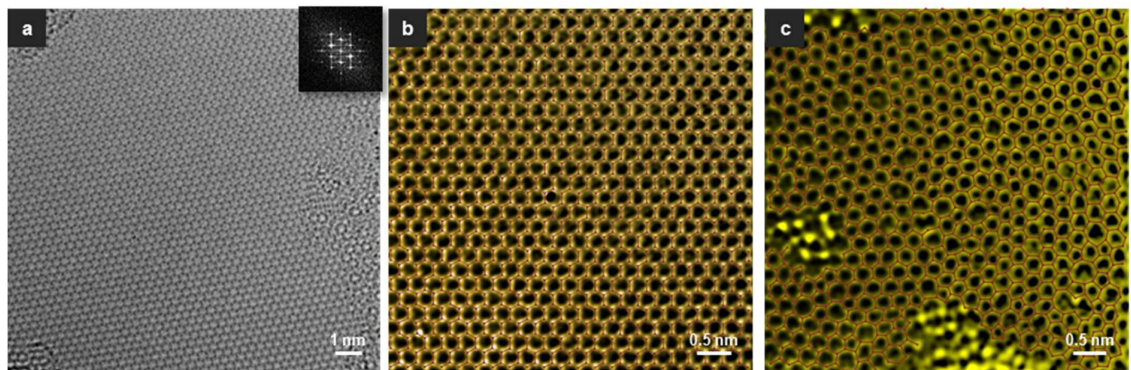

**Figure S8 Atomic structure of graphene and MAC. a**, monochromated HRTEM image of graphene and the corresponding Fourier transform (top right inset). **b**, large scale atom-by-atom mapping from within (**a**). The contrast of the image is inverted and false coloured for better visibility. **c**, large scale atom-by-atom mapping of MAC from Fig. 2b without the colour overlay.

We have drawn up boundaries for MAC with the upper bound where crystallites are not separated by atomically sharp grain boundaries but by continuous random network (CRN) regions of at least 3 carbon atoms wide and with the lower bound where crystallites exist with size of at least a hexagonal ring surrounded by six hexagonal rings. The differences are seen between strained crystallites and nanocrystalline graphene with 1-3 nm grain size and sharp grain boundaries. Selective area electron diffraction (SAED) patterns of the samples in Figure 2h, i (Fig. S9), confirms the amorphous nature of MAC by the characteristic diffuse halo (Fig. 2j), in contrast to sharp first and second order diffraction rings for nanocrystalline graphene (Fig. 2k). Dark-field TEM (DF-TEM) also reveals this, with only nanocrystalline graphene showing crystallinity (Extended Data Fig. 2). We note that crystallites are also observed in 2D ionic systems, e.g. in amorphous $SiO_2$ bilayer films grown on Ru[16] and graphene[34], although these amorphous regions are only continuous up to ~10 nm and needs to be supported by their growth substrates for stability.

A critical requirement for MAC formation is the low synthesis temperature. From TEM data, MAC grown with a process temperature between 200 °C and 500 °C are observed to be nearly identical, indicative of the robustness of the present growth method. Beyond 500 °C, the synthesis process results in nanocrystalline graphene and polycrystalline graphene formation (Fig. S10, to be published).

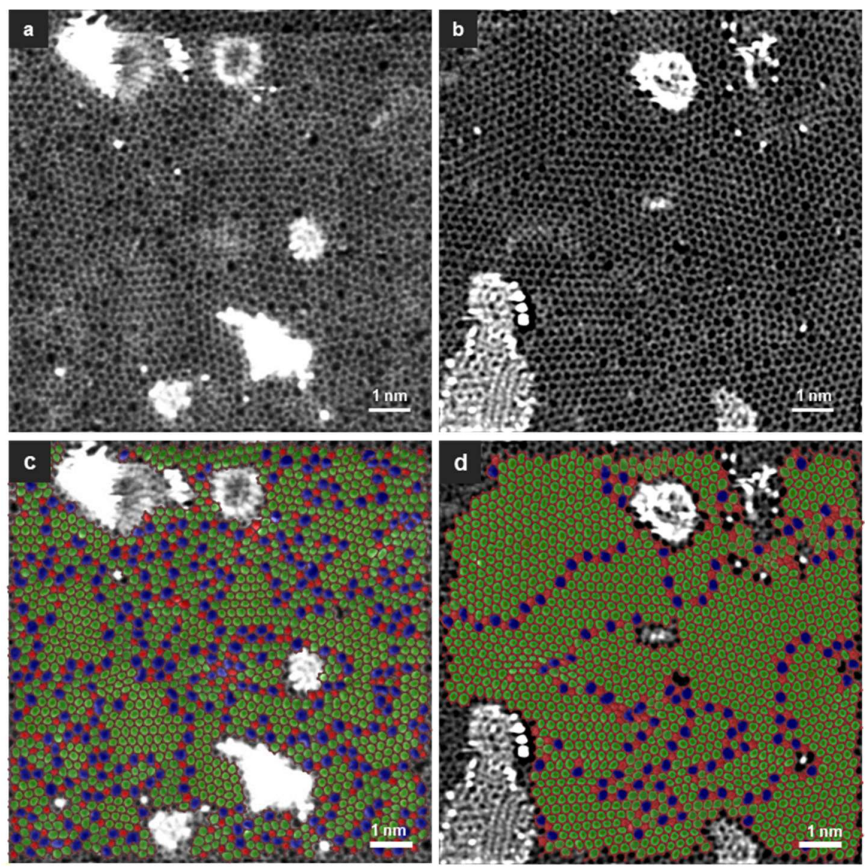

**Figure S9** Atomic structure of **a, c,** MAC and **b, d,** nanocrystalline graphene. **a-b**, captured by STEM without the false colour overlay (in Fig. 2h, i) and **c-d**, with false colour overlay with hexagons (green), pentagons (red) and heptagons/octagons (blue) shaded.

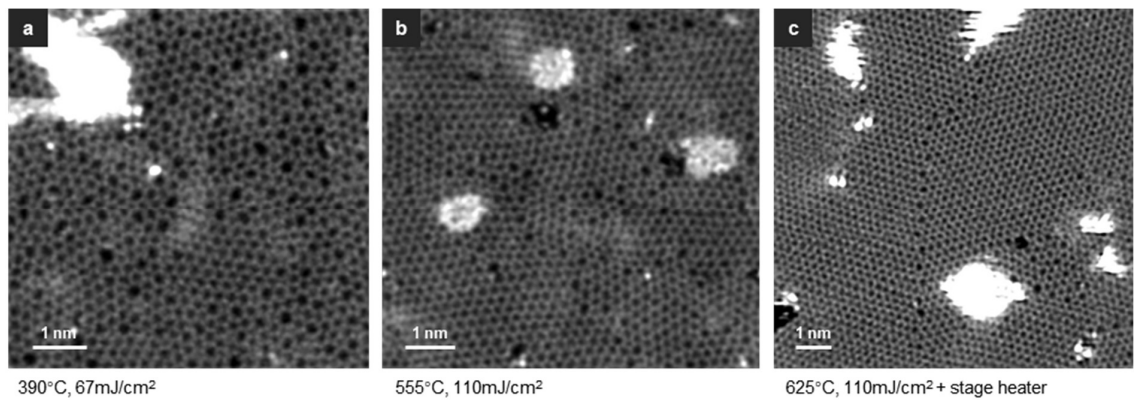

390°C, 67mJ/cm²     555°C, 110mJ/cm²     625°C, 110mJ/cm² + stage heater

**Figure S10** Comparison between the STEM images of different crystallinity. **a**, MAC, **b**, nanocrystalline graphene and **c**, polycrystalline graphene. Samples were grown by controlling the laser intensity and hence the growth temperature as shown. The stage heater was used together with the laser for polycrystalline graphene growth in (**c**).

MAC is also distinct from defective graphene created by electron beam irradiation[9]. Albeit only very small areas, these materials have a statistically inhomogeneous structure and include an unavoidable density of holes that reflects local damage patterns by carbon loss. This feature is indeed evident at intermediate irradiation levels when crystallites are present with a mix of 5-, 6-, 7-, and 8-member rings, but the crystallites are not randomly oriented as is necessary for a truly amorphous structure. Beyond the intermediate range, when crystallites are not present, the material is probably best characterized as heavily damaged graphene and is not representative of truly monolayer amorphous carbon. Such structures are also distinctly different from a synthesised amorphous structure and suggest that their crystallinity can in principle be restored by sequential removal of defects[35]. Nevertheless, these studies have been valuable in that they establish the resilience of threefold-coordinated bonding of carbon atoms in 2D and the effect of non-hexagonal bonding patterns on physical properties.

**Supplementary Note 4: Fracture Toughness of MAC**

MAC membrane undergoes plastic deformation with applied force, leading to force vs membrane deflection curves to change with increase indentation force as seen in Fig. 3d. $E_{2D}$ is therefore consistently calculated from a reproducible curve after multiple indentations at 100 nN indentation force. The correlation between pre-tension and $E_{2D}$ is observed, where stiffness increases under tension (Fig. S11a, b). Plastic deformation is especially significant directly under the indentation tip where strain is highest. In fact, we observed a peak at the centre of the MAC membrane, distinct from graphene which elastically stretches and reverts to a flat membrane. Furthermore, significant rippling is observed over the suspended MAC, but plastic deformation allows for flattening of the membrane under applied force (Fig. S11c, d). MAC is therefore a soft material and obtains a stiff pre-stretched structure[36] after plastic deformation. For crystalline 2D materials, significant plastic deformation has not been observed. Also, in MAC, indentation rupture is restricted and does not propagate, as compared to graphene where rapid crack propagation lead to entire membrane collapse (Fig. S11e). Both mechanical properties indicate a high fracture toughness for MAC.

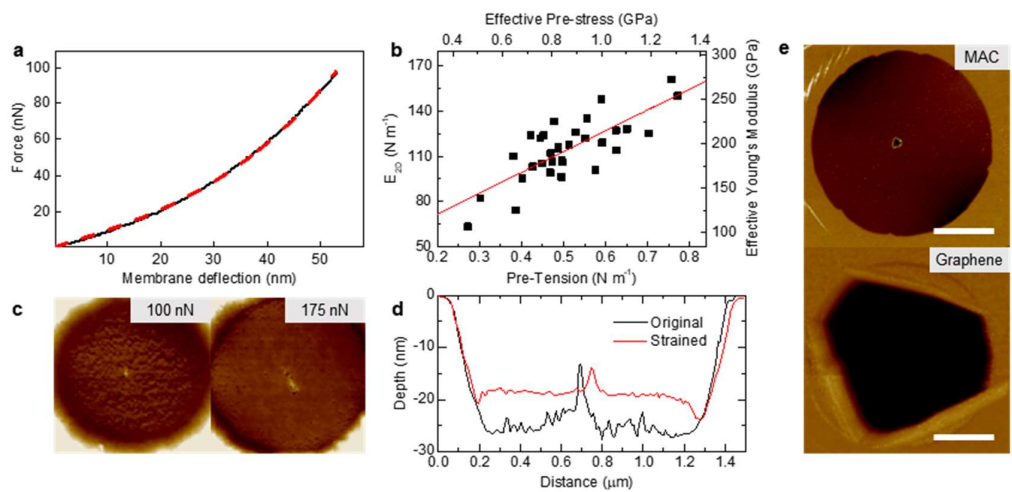

**Figure S11 Indentation measurements on MAC. a**, force vs deflection curve and curve fitting to Equation (1) for the calculation of $E_{2D}$ and pre-tension values. Curved used in the fitting were taken only after multiple indentation at 100 nN leads to a reproducible curve. **b**, Correlation between 2D elastic modulus vs pre-tension with linear fit (red line, intercept = 44±10, slope = 138±20). **c**, (left) AFM topography of suspended MAC on 1 μm diameter wells have visible rippling (100 nN indentation force applied to depress MAC into the well). (right)

After measurements at high indentation force, these samples have significantly reduced roughness (175 nN indentation force applied). **d**, AFM height profile across the centre of (**c**). **e**, when breaking force is applied to MAC, a hole is seen in the membrane. This is a stark contrast to graphene, where the entire membrane collapses catastrophically. Scale bar, 1 μm (**e**) for both MAC and graphene image.

**Supplementary Table**

| # | Type | Length, $L$ (nm) | $W/L$ ratio | Power, N | Offset, $\beta$ |
|---|---|---|---|---|---|
| TL0 | 3L | 200 | 250 | -5.5 | 24.3 |
| TL1 | 3L | 200 | 250 | -6.3 | 25.1 |
| TL2 | 3L | 500 | 100 | -6.5 | 25.1 |
| TL3 | 3L | 200 | 250 | -4.9 | 21.7 |
| TL4 | 3L | 300 | 166 | -6.5 | 25.3 |
| BL1 | 2L | 200 | 250 | -7.5 | 28.4 |
| BL2 | 2L | 300 | 166 | -4.6 | 20.5 |
| BL3 | 2L | 500 | 100 | -4.4 | 19.8 |
| ML1 | 1L | 1000 | 50 | -6.5 | 26.9 |
| ML2 | 1L | 1000 | 50 | -7.6 | 30 |

**Table S1 Dataset for temperature dependence of MAC.** The list of the samples in Extended Data Fig. 5 with geometrical sizes and the parameters of the linear fits, where the proportionality constant, $\alpha = 10^{\beta}$.

**Supplementary References**